\documentclass{statsoc}

\usepackage[a4paper]{geometry}
\usepackage{graphicx}
\usepackage[textwidth=8em,textsize=small]{todonotes}
\usepackage{amsmath}
\usepackage{natbib}
\usepackage{hyperref}
\usepackage{float}
\restylefloat{table}

\newcommand{\set}[1]{{\{ #1\}}}
\newcommand{\bigset}[1]{{ \big \{ #1 \big\}}}
\newcommand{\biggset}[1]{{ \bigg \{ #1 \bigg\}}}
\newcommand{\bb}[1]{\boldsymbol{ #1 }}
\newcommand{\given}{\, | \,}
\newcommand{\textspace}[2] {{\hspace{#2} \textrm{ #1 } \hspace{#2}}}

\title[Spatially Selected and Dependent Random Effects for SAE]{Spatially Selected and Dependent Random Effects for \\ Small Area Estimation with Application to Rent Burden}
\author{Sho Kawano, Paul A. Parker, Zehang Richard Li}
\address{University of California, Santa Cruz, USA. }
\address{Email: shkawano@ucsc.edu}

\begin{document}
\begin{abstract}
Area-level models for small area estimation typically rely on areal random effects to shrink design-based direct estimates towards a model-based predictor. Incorporating the spatial dependence of the random effects into these models can further improve the estimates when there are not enough covariates to fully account for spatial dependence of the areal means. A number of recent works have investigated models that include random effects for only a subset of areas, in order to improve the precision of estimates. However, such models do not readily handle spatial dependence. In this paper, we introduce a model that accounts for spatial dependence in both the random effects as well as the latent process that selects the effects. We show how this model can significantly improve predictive accuracy via an empirical simulation study based on data from the American Community Survey, and illustrate its properties via an application to estimate county-level median rent burden. \\

\end{abstract}

\section{Introduction}\label{sec:intro}

Small Area Estimation (SAE) has become an integral tool in official statistics to produce estimates for quantities of interest, such as means and totals, for subpopulations with small sample size. The term \textit{small area} usually refer to geographic subregions but can also be demographic groups or a combination of the two. SAE models play a vital role in the effective implementation of government policies. One of the many policy-relevant estimates of interest is rent burden, defined as the share of household income spent on rent. 

Cost of housing is a very pressing issue in the United States. According to 2017-2021 ACS 5-year estimates, over 19 million U.S. renter households were paying more than 30\% of income on rent, making them \textit{cost burdened} \citep{cromwell_renters_2022}. More recently in 2022, 22.5 million households, half of all U.S. renters, were estimated to be cost burdened, based on the ACS 2022 1-year estimates \citep{us_census_bureau_american_2023, joint_center_for_housing_studies_of_harvard_university_americas_2024}. Furthermore, 12.1 million households are estimated to be \textit{severely cost burdened}, paying more than half of their incomes on rent.  These severe rent burdens heavily affect households with lower incomes that do not have sufficient funds to cover basic necessities after paying rent \citep{us_bureau_of_labor_statistics_consumer_1914, desmond_heavy_2018}. 

U.S. housing policy is highly decentralized. Although the central (or `federal') government administers subsidies to housing consumers, most subsidies are directed at homeowners rather than renters, despite renters typically falling into lower income categories\citep{crump_us_2021}. In other key areas of housing policy, especially in land use planning and development permitting, the federal government has quite limited powers \citep{quigley_housing_2008}. Most of the consequential policy debates affecting renters today are therefore happening at the level of local and state governments. Thus, providing accurate and granular estimates at the local and state level is critical to good policy making.  

Direct survey estimators often fail to produce reliable estimates at the level of a small area (e.g. counties or census tracts). This is due to the fact that transnational surveys like the American Community Survey are usually designed for accuracy at a high level of aggregation. Thus, model-based approaches may become necessary for reliable small area estimation without development of new surveys. In this paper, we primarily focus on what is known as area-level modeling, where aggregated direct estimates at discrete areas of interest are modeled. This approach contrasts with unit-level small area estimation models, where individual survey responses are directly modeled. For a review of unit-level models, see \cite{parker_comprehensive_2023}.

The foundational area-level model was introduced by \citet{fay_estimates_1979}. Their model utilizes a regression-based estimate with independent, normally distributed random effects that have zero mean and a common variance. The estimates from such a model can be viewed as a weighted combination of the direct survey estimates and the regression-based estimates, where the weights are constructed according to the ratio of the survey variance and the variance of the random effects.  If the survey variance is relatively large, then the Fay-Herriot (FH) model puts more weight on the regression based estimate and vice versa. This allows for more accurate estimation in areas with small sample size by \textit{borrowing strength} from areas with larger samples. The FH model has become one of the most widely used tools for SAE. Extensions that generalize the FH model to new types of data and/or relax modeling assumptions have grown substantially over the years. 

One convenient assumption the FH model makes is that the survey variances are fixed and known, when in reality they are estimated via the sample design and then plugged into the model.   \citet{you_hierarchical_2016} and  \citet{sugasawa_bayesian_2017} introduce Bayesian models that use a weighted estimate similar to the Fay-Herriot approach to model the survey variance in addition to the small area mean, with the latter allowing the use of covariates. These models were compared in \citet{you_small_2021}.  \citet{parker_conjugate_2023}  further extend the approach  by proposing a model that yields conjugate full conditional distributions to further allow for spatial modeling of the survey variances. 

There is also a growing body of literature dedicated to proposing alternate structures for the random effects used within the FH model. For instance, instead of assuming independence, one might consider a spatial structure to the random effects. This allows for models that account for spatial dependence of the quantities of interest, which is commonly observed in many SAE applications. A common distributional choice is the conditional autoregressive (CAR) structure. Examples of previous work using the CAR structure include \citet{zhou_hierarchical_2008} and  \citet{porter_small_2015}, the latter providing an extension for multivariate data. Another common choice is the simultaneous autoregressive (SAR) structure. For examples, see \citet{singh_spatio-temporal_2005}, \citet{petrucci_small_2006}, and  \citet{schmid_spatial_2014} .  The quickly growing spatial SAE literature suggests that incorporating spatial dependence can substantially improve the precision of model-based estimates. 

Another area of exploration has been to address the normality and common variance assumptions of the random effects. These assumptions can cause the FH model to be less robust in situations where large random effect values are needed for relatively few areas. Many recent papers have proposed extensions that move away from one or both of these simplifying assumptions. An overview of the this topic is given by \citet{jiang_robust_2020}.  One approach is to assume heavier-tailed distributions for the random effects, which was explored by \citet{datta_robust_1995} and \citet{fabrizi_robust_2010}, among others.   \citet{chakraborty_two-component_2016} use a mixture of two normal distributions, one with a small variance and the other with a large variance, to relax the common variance assumption. Finally, \citet{janicki_bayesian_2022} take a novel approach by using a Bayesian nonparametric (BNP) model to estimate counts for multi-way contingency tables (e.g. race and age groups by county). Using a BNP prior to model both the fixed and random effects is an effective way to model contingency tables that exhibit heterogeneity both in the coefficient-response relationships as well as spatial patterns.

A more recent (and somewhat related) avenue of research was spurred by a number of works that have questioned whether all of the random effects are necessary in the presence of appropriate covariates \citep{datta_model_2011}. This has led to development of models that use \textit{shrinkage} priors, pushing the unnecessary random effects towards zero.  More specifically, \citet{datta_small_2015} proposed an SAE model with a discrete-normal spike-and-slab prior that shrinks the unnecessary random effects to zero. A continuous analog to the Datta-Mandal model were introduced by \citet{tang_modeling_2018} with a multivariate extension by \citet{ghosh_multivariate_2022}, using the class of global-local shrinkage priors for the random effects. In the global-local approach, the variance of each random effect is a product of a global shrinkage parameter and a local parameter that serves as an area-specific adjustment. 

Shrinkage guided by the spatial structure of the areas of interest may further improve accuracy of model-based estimates. Intuitively, if a random effect is not necessary in a given region, it may be likely that it is also not necessary in a neighboring area. On the other hand, if a random effects are necessary for two neighboring areas, the value of the random effects may be similar as well. Therefore, spatial dependence can be built in on two different levels: in the latent selection process that selects/shrinks the random effects as well as in the magnitude of the random effects that are selected. 

Building in spatial dependence for the random effects themselves is somewhat straightforward within the global-local framework, as demonstrated by \citet{tang_global-local_2023}, who used a CAR prior for the random effects.  However, building in spatial dependence for the shrinkage process is complicated by the distributional requirements of the global-local prior. For instance, using a spatial horseshoe prior on the random effects would require the use of a Gaussian copula on the local parameters \citep{tadesse_bayesian_2021}. Given these complications, a modeling approach that uses a spike-and-slab prior may provide a more natural starting point.  

In this paper, we introduce a model that incorporates spatial dependence on both the selection process and the random effects themselves, extending the work of \citet{datta_small_2015}. The remainder of this paper is organized as follows. In Section \ref{sec:basics}, we first discuss area-level modeling, introducing the Fay-Herriot model in addition to the relevant spatial and shrinkage extensions. In Section \ref{sec:model}, we introduce our proposed model and discuss prior specification. In Section \ref{sec:post_inference}, we describe posterior inference for our proposed model, including the full conditional distributions of each parameter. In Section \ref{sec:sim_study} we conduct an empirical simulation study that shows how the proposed model can significantly improve the accuracy of both point and interval estimates of median rent burden, using ACS county-level data for North Carolina. In Section \ref{sec:data_analysis} we use the proposed model to estimate median rent burden for all counties in the South Atlantic Census Division. Finally, we conclude with a discussion in Section \ref{sec:conclusion}. 

\section{Area-Level Modeling} \label{sec:basics}

Consider a survey in a study region partitioned into $n$ small areas. Let the small area means  $\bb{\theta} = (\theta_1, \dots, \theta_n)$ be the quantities of interest. For each area $i=1, \dots, n$, a survey is used to provide design-based direct estimates of $\theta_i$, denoted as $y_i$, with survey variances $d_i$ which are assumed to be known.  Let $N_i$ represent the population size in area $i$ and $y_{ij}$ denote the  value of some variable of interest for the $j$th population unit in area $i$, where $j \in \set{1, \cdots, N_i}$.  Under the design-based perspective, the small area means are treated as fixed but unknown quantities (i.e. $\theta_i = \tfrac{1}{N_i}\sum_{j=1}^{N_i} y_{ij}$).  The commonly used \citet{horvitz_generalization_1952} estimator incorporates design information through inverse-probability weighting, 
\begin{align*}
  y_i = \tfrac{1}{N_i} \sum_{j \, \in \, \mathcal{S}_i}  \frac{y_{ij}}{\pi_{ij}}
\end{align*}
where $\mathcal{S}_i$ denotes the collection of sample indices for area $i$ and $\pi_{ij}$ represents the probability that the $j$th unit in area $i$ is included in the sample. Direct estimators like the Horvitz-Thompson have important properties such as being design unbiased. 

However, for areas where the sample size is small, direct estimators can have unreasonably high standard errors, which may necessitate the use of a model. A standard area-level model can be written as
\begin{align*}
  [\bb{y} \given \bb{\theta}] \sim N_n(\bb{\theta}, \bb{D}) \textspace{and}{1mm} \bb{\theta} = \bb{X} \bb{\beta} + \bb{u}, 
\end{align*}
where $\bb{y} = (y_1, \dots, y_n)^\top$, the $n \times n$ covariance matrix is $\bb{D}=\textrm{diag} \set{d_1, \cdots, d_n}$, $\bb{X}$ is a $n \times j$ full-rank covariate matrix, and $\bb{\beta}$ is the corresponding coefficient vector (a.k.a. fixed effects). Finally, the length $n$ vector $\bb u$ represents the random effects. The random effects $\bb u$ capture the variability in $\theta_i$ that cannot be fully explained by the covariates in $\bb{X}$. A great deal of research in SAE is focused on the distributional choice for $\bb u$. 

The random effects for the foundational Fay-Herriot model \citep{fay_estimates_1979}  are assumed to be independent and identically distributed (IID) normal random variables with common variance: 
\begin{align*}
  [\bb u &\given \sigma_{FH}^2] \sim N_n(\bb 0, \sigma_{FH}^2 \bb{I}).
\end{align*}
In a Bayesian setting, the Fay-Herriot model is often used with the improper prior  $\pi(\beta, \sigma_{FH}^2) \propto 1$. Despite its widespread acceptance, there are some shortcomings with this model, as previously mentioned. These limitations are generally driven by the assumption that the random effects are IID and normally distributed.  


\subsection{Incorporating Spatial Dependence in the Random Effects}
Instead of assuming independence among random effects, one could model some dependence structure in the random effects
\begin{align*}
  [\bb u \given \sigma_{S}^2, \rho] \sim N_n(\bb 0, \sigma_{S}^2 \, \bb{Q}^{-1}{(\rho)}),
\end{align*}
where $\sigma^2_S$ represents the random effect variance and $\rho$ denotes additional parameters needed for $\bb Q$, the precision matrix. A common choice for $\bb Q$ is the Conditional Autoregressive (CAR) structure \citep{besag_spatial_1974}. The CAR precision matrix has the form $\bb{Q}= diag\{A_{ii}\}^n_{i=1} - \rho \bb{A}$ where $\rho \in (-1, 1)$ is the spatial correlation parameter and $\bb A$ is a $n \times n$ spatial adjacency matrix whose $i,k$th element $a_{ik}$ is equal to one if areas $i$ and $k$ share a border, and equal to zero otherwise. Positive (negative) values of $\rho$ indicate positive (negative) spatial correlation with stronger spatial correlation indicated by values close to $\pm 1$, although it is common practice to restrict $\rho \ge 0$.  Setting $\rho=1$ results in the degenerate Intrinsic Conditional Autoregressive (ICAR) prior. 

Existing literature indicates that the CAR model may lead to misleading results when there is no spatial correlation actually present in the data \citep{leroux_estimation_2000, wakefield_disease_2007}. For this reason, one may instead consider a Besag York Mollié (BYM) prior  \citep{besag_bayesian_1991} to model the random effects using both spatial and non-spatial components,
\begin{align*}
  \bb u = \bb v_1 +  \bb v_2  &\textrm{ where } [\bb v_1] \sim N_n(\bb 0, \sigma_{1}^2 \, \bb{I}) \textspace{and}{2mm} [\bb v_2] \sim N_n(\bb 0, \sigma_{2}^2 \, \bb{Q}^-)
\end{align*}
where sum-to-zero constraints are placed on $\bb v_1, \bb v_2$ for identifiability. Note that $\bb{Q}^-$ is the generalized inverse of the ICAR precision matrix.  Under this prior, the random effects are a sum of two components: IID normal effects $\bb v_1$ and the ICAR spatial effects $\bb v_2$. 

In order to aid in prior specification of $\sigma_{2}^2$, we found it useful to scale the ICAR precision matrix by dividing it by the geometric mean of the diagonal elements, as recommended by  \citet{sorbye_scaling_2014}.   Scaling the ICAR precision matrix allows the parameter $\sigma_2^2$ to reflect the typical variance, allowing the prior of the IID variance $\sigma^2_1$ to be roughly comparable to that of $\sigma^2_2$ \citep{riebler_intuitive_2016, freni-sterrantino_note_2017}. 

\subsection{Incorporating Random Effect Shrinkage} 
 \citet{datta_small_2015} introduced the first SAE model that uses Bayesian variable selection methodology for the random effects. Specifically, they use a spike-and-slab prior that assumes a discrete-normal mixture for the random effects
\begin{align*}
  \bb u &= \bb \delta \odot \bb v \\  [v_i \given \delta_i=1, \, \sigma_{DM}^2] \, & \stackrel{ind}\sim \, N(0, \sigma_{DM}^2), \\  [v_i \given \delta_i  =0] & = 0 \\ 
  [\delta_i | p]  & \stackrel{iid}\sim Bernoulli(p) \\  [p] &\sim Beta(a, b).
\end{align*}
To complete the model, they use an inverse-gamma prior on $\sigma^2_{DM}$ and $\pi(\beta) \propto 1$. Note that we use $\odot$ to represent the operator for an element-wise vector product.  In the Datta-Mandal (DM) model, a given random effect, $v_i$, for area $i$ is only included if the corresponding selection indicator, $\delta_i$, is equal to 1. This approach relaxes the common variance assumption because $\sigma_{DM}^2$ only applies to random effects for areas that are included (i.e. $\delta_i=1$). Meanwhile, the effects that are not included are degenerate with zero variance. 

The mechanism behind the shrinkage of the random effects for the DM model can be understood from the posterior selection probability, $\check p_{i}$, for a given area $i$, 
\begin{align}\label{eq:iprob1}
  \check p_i = \frac{p \cdot \phi \big(y_i \given x_i^\top \bb{\beta} + v_i, \, d_i \big)}{p \cdot \phi \big(y_i \given x_i^\top \bb{\beta} + v_i, \, d_i \big) + (1-p) \cdot \phi \big(y_i \given x_i^\top \bb{\beta}, \, d_i \big)},
\end{align}
where $\phi (z \given \mu, \kappa^2)$ is the normal density function with mean $\mu$ and variance $\kappa^2$, evaluated at the point $z$.  The value of $\check p_i$ can be interpreted as the posterior probability that a random effect is necessary in area $i$. The global level of shrinkage across all areas is influenced by the parameter $p$.  If the posterior distribution of $p$ is close to one, it indicates that most areas need random effects and vice versa. Area-specific selection probabilities also depend on the normal likelihood evaluations both with and without the random effect. For example, even if $p$ has mass towards zero, if the fit without the random effect is very poor for a given area $i$, then $\phi \big(y_i \given x_i^\top \bb{\beta} + v_i, \, d_i \big)$ will be much greater than $\phi \big(y_i \given x_i^\top \bb{\beta}, d_i \big)$, which will be reflected in the selection probability. 

\section{The Spatially Selected and Dependent Random Effects Model} \label{sec:model}

We propose a model that accounts for spatial dependence on both the selection process and the random effects themselves. The data model is given by
\begin{align*} 
  [\bb y \given \bb{\theta}] \sim N_n(\bb{\theta}, \bb{D}) \textspace{and}{1mm} \bb{\theta} = \bb{X} \bb{\beta} + \bb{u},
\end{align*}
where $n$ again is the number of small areas. Note that here, the direct estimates, $\bb y$, are scaled to have zero mean, and unit variance, and the survey variances are correspondingly scaled to form the diagonal matrix $\bb{D}$. The covariate matrix, $\bb X$, is full-rank. Similar to \cite{datta_small_2015}, we consider a spike-and-slab prior for the random effects, although we include spatial structure, 
\begin{align*}
  \bb u &= \bb \delta \odot (\bb v_1 +\bb v_2) \\  [\bb v_1] & \sim N_n(\bb 0, \sigma_{1}^2 \, \bb{I}) \\ [\bb v_2] & \sim N_n(\bb 0, \sigma_{2}^2 \, \bb{Q}^-)\\ 
    [\delta_i|p_i] & \stackrel{ind}\sim Bernoulli(p_i).
\end{align*}
Note that for identifiability purposes, we place a sum-to-zero constraint on $\bb v_1$ and  $\bb v_2$, while $\bb{Q}^-$ is the generalized inverse of the ICAR precision matrix, which is scaled by the geometric mean of the diagonal elements, as suggested by \citet{sorbye_scaling_2014}.  Also note that the selection indicators themselves are assumed independent, but each have a separate selection probability parameter, $p_i$.  We further model dependence within the selection process through the prior on these selection probabilities:
\begin{align*}
    logit(\bb p) &= \bb \psi_1 + \bb \psi_2 \\ [\bb \psi_1|s_1^2] & \sim N_n(\bb 0, s_{1}^2 \, \bb{I}) \\ [\bb \psi_2|s_2^2] & \sim N_n(\bb 0, s_{2}^2 \, \bb{Q}^-),
\end{align*}
where, again, $\bb{Q}^-$ is  the generalized inverse of the scaled ICAR precision matrix. Note that our model uses the BYM structure on both the random effects as well as the logit effects that form the selection probabilities. This allows spatial and non-spatial components in both the selection process and the random effects themselves. We refer to this new model as the Spatially Selected and Dependent (SSD) random effects model. 

\subsection{Prior Specification}

To simplify prior specification for the SSD model, we recommend fitting on scaled data $(\bb{y}, \bb{D})$, similar to what is done in other shrinkage methods such as the Bayesian Lasso \citep{park_bayesian_2008}.  For the random effect variance parameters, we specify inverse-gamma priors $[\sigma_1^2] \sim IG(c, d)$, and $[\sigma_2^2] \sim IG(c, d)$ where $c, d$  are the shape and scale hyperparameters, respectively.  The choice of prior for $\sigma_1^2$ and $\sigma_2^2$ can be important, especially since these parameters can influence the level of shrinkage. In order for the model to be able to distinguish between the degenerate \textit{spike} at zero and the \textit{slab} normal distribution, a prior needs to avoid variance values that are too small, while still being sufficiently diffuse. Thus, for fitting on scaled data, we recommend a prior such as $IG(5, 5)$, which, in our experience, provided the necessary qualities. For the regression coefficients, we specify the proper prior $[\bb \beta] \sim N_j (\bb 0, k^2  \bb{I})$ where the hyperparameter $k^2$ is sufficiently large as to be non-informative. This could depend on the scale of the data, but for the suggested standardized data, we recommend $k^2=100^2$. Lastly we let $s_{1}^2 \sim IG(a, b)$ and  $s_{2}^2 \sim IG(a, b)$ again for fixed hyperparameters $a, b$, where we recommend $a=5$ and $b=10.$

\section{Posterior Inference}  \label{sec:post_inference}

Let $\bb \Omega$ denote the set of parameters in the SSD model. Then, the full posterior distribution can be written, up to a constant of proportionality, as 
\begin{align*}
  \pi(\bb \Omega | \bb{y}, \bb{X}, \bb{D}) &\propto exp \biggset{-\tfrac{1}{2} (\bb{y} - \bb{X} \bb{\beta} - \bb{\delta} \odot [\bb{v_1} + \bb{v_2}])^\top \bb{D}^{ -1} (\bb{y} - \bb{X} \bb{\beta} - \bb{\delta} \odot [\bb{v_1} + \bb{v_2}]) } \\
  &\times   (\tfrac{1}{\sigma_1^2} )^{n/2}  (\tfrac{1}{\sigma_2^2} )^{n/2}  exp\bigset{-\tfrac{1}{2 \sigma_1^2} \bb{v_1}^\top\bb{v_1} -\tfrac{1}{2 \sigma_2^2} \bb{v_2}^\top \bb{Q}\bb{v_2} } \\
  &\times \prod_{i=1}^n p_i^{\delta_i} (1-p_i)^{1-\delta_i}  \\
  &\times (\tfrac{1}{s_1^2} )^{n/2}  (\tfrac{1}{s_2^2} )^{n/2}  exp\bigset{-\tfrac{1}{2 s_1^2} \bb{\psi_1}^\top\bb{\psi_1} -\tfrac{1}{2 s_2^2} \bb{\psi_2}^\top \bb{Q}\bb{\psi_2} } \\
  &\times \pi(\beta) \, \pi(\sigma^2_1) \, \pi(\sigma^2_2) \, \pi(s_1^2) \, \pi(s_2^2).
\end{align*}
  This model leads to full conditional distributions all from standard parametric families, when Pólya-Gamma Data Augmentation is applied \citep{polson_bayesian_2013}. Then, Gibbs sampling can be used to sample from the posterior distribution.  We will denote $\bb{y}, \bb{X}, \bb{D}$ as $data$ for ease of notation.

\paragraph{Full conditional distributions of $\bb{v_1}$, $\bb{v_2}$,  and $\bb{\beta}$.}
The fixed and random effects $\bb \beta$ and $\bb{v_1, v_2}$ can be sampled in a block.  In order to do so, we set $\bb \gamma = \begin{pmatrix} \bb{\beta}^\top, &   \bb{v}_1^\top, & \bb{v}_2^\top \end{pmatrix}^\top$ and  $\bb{Z} = \begin{pmatrix} \bb{X} & \Delta & \Delta \end{pmatrix}$  where $\Delta = diag\{\delta_i\}^n_{i=1}$ and  $\bb{Z} \gamma = \bb{X}  \bb{\beta} + \bb{\delta} \odot (\bb{v_1 + v_2})$.  Provided that there exists $i$ such that $\delta_i \neq 0$, the full conditional of $\bb \gamma$ is then 
\begin{align*}
    [\bb \gamma \given \sigma_1^2, \sigma_2^2, \bb \delta, data] \sim N_{j + 2n} \bigg( \bb{P}_\gamma^{ -1} \bb{D}^{ -1} \bb{Z}^{\top }   \bb{y}, \, \bb{P}_\gamma^{ -1}\bigg )
\end{align*}
where
\begin{align*}
   \bb{P}_\gamma = \bb{Z}^\top \bb{D}^{ -1} \bb{Z} + \bb{\Lambda}_\gamma  \textspace{and}{1mm} \bb{\Lambda}_\gamma =  \begin{pmatrix}
     \bb{I}_j/ k^2  & 0 & 0 \\
     0 &  \bb{I}_n / \sigma^2_1 & 0 \\
    0 & 0 &  \bb{Q} / \sigma^2_{2}
\end{pmatrix}. 
\end{align*}
Note that $\bb{I}_m$ denotes an identity matrix of rank $m$ and $\bb{Q}$ again is the scaled ICAR precision matrix. If $\delta_i=0$ for all $i \in \set{1, \cdots, n}$, then $\bb v_1  = \bb v_2 = \bb 0$ and the posterior distribution of $\bb \beta$ is: 
\begin{align*}
    [\bb \beta \given \sigma_1^2, \sigma_2^2, data] \sim N_{j} \bigg( \bb{P}_\beta^{ -1} \bb{D}^{ -1} \bb{X}^{\top }   \bb{y}, \, \bb{P}_\beta^{ -1}\bigg ) \textspace{where}{2mm} \bb{P}_\beta = \bb{X}^\top \bb{D}^{ -1} \bb{X} + \bb{I}_j/ k^2.
\end{align*}

\paragraph{Full conditional distribution of $\bb{\delta}$.} 
For the SSD model, the posterior selection probability for area $i$ is 
\begin{align}\label{eq:iprob2}
  \tilde p_{i} &= \frac{p_i \cdot \phi \big(y_i \given x_i^\top \bb{\beta} + v_{1i}+v_{2i}, \, d_i \big)}{p_i \cdot \phi \big(y_i \given x_i^\top \bb{\beta} +v_{1i}+v_{2i}, \, d_i \big) + (1-p_i) \cdot \phi \big( y_i \given x_i^\top \bb{\beta}, \, d_i \big)} \\
  \, \nonumber \\
  &= p_i \, \big / \bigg ( p_i + (1-p_i) exp \bigset{\tfrac{1}{2 d_i}(v_{1i}+v_{2i})^2 - (y_i-x_i^\top \bb{\beta})(v_{1i}+v_{2i})}\bigg ), \nonumber
\end{align}
where $y_i$ are the scaled direct estimates, and $d_i$ are the scaled survey variances. Note that the structure of the selection probability here is very similar to (\ref{eq:iprob1}) . The full conditional distribution of $\bb \delta$ is then 
\begin{align*}
    [\delta_i \given p_i, v_{1i}, v_{2i}, \bb \beta, data ]  \stackrel{ind}\sim  Bern (\tilde p_i) \textspace{for}{1mm} i = 1, \cdots, n.
\end{align*}

\paragraph{Full conditional distributions of $\bb \psi_1$, $\bb \psi_2$, and latent variables $\bb w$.}
The posterior conditional distribution of  $\bb \psi_1, \bb \psi_2$ can be sampled with Pólya-Gamma data augmentation \citep{polson_bayesian_2013}. The core integral identity of this approach is the following:
$$\frac{(e^\varphi)^a}{(1+e^\varphi)^b} = 2^{-b} e^{\kappa \varphi} \int^\infty_0 e^{-w \varphi^2/2} p(w) dw$$
where $\kappa = a - b/2$ and $w$ has a Pólya-Gamma (PG) distribution, $PG(b, 0)$.  In our model, the probability mass function for $\delta_i$ is
$$p_i^{\delta_i} (1-p_i)^{1-\delta_i} = \bigg ( \frac{e^{\varphi_i}}{1+e^{\varphi_i}} \bigg )^{\delta_i} \bigg ( \frac{1}{1+e^{\varphi_i}} \bigg )^{1-\delta_i}= \frac{(e^{\varphi_i})^{\delta_i}}{(1+e^{\varphi_i})}$$
where $\varphi_i = \psi_{1i} + \psi_{2i} = logit(p_i)$. Setting $a=\delta_i$ and $b=1$ we use the PG data augmentation strategy and sample the full conditional of $\bb \psi_1, \bb \psi_2$  in a block.  We set the blocked logit effects  $\mathit{\Psi}  =\begin{pmatrix}
    \bb \psi_1^\top, & \bb \psi_2^\top
\end{pmatrix}^\top$ and $\bb{H} = \begin{pmatrix} \bb{I}_n & \bb{I}_n \end{pmatrix}$.  We introduce PG latent variables $[w_i] \sim PG(1, 0)$ and set $\bb W = diag\{w_i\}^n_{i=1}$. Also, we set $\bb \kappa = \begin{pmatrix}
    \delta_1 - 1/2, & \cdots, & \delta_n - 1/2
\end{pmatrix}^\top$. The blocked logit effects can be sampled from
\begin{align*}
    [\mathit{\Psi} \given s_1^2, s_2^2, \bb W] \sim N_{2n} \bigg( \bb{P}_\Psi^{ -1} \bb{H}^\top \bb \kappa, \, \bb{P}_\Psi^{ -1}\bigg )
\end{align*}
where
\begin{align*}
   \bb{P}_\Psi = \bb{H}^\top \bb{W} \bb{H} + \bb{\Lambda}_\Psi \textspace{and}{1mm} \bb{\Lambda}_\Psi =  \begin{pmatrix}
     \bb{I}_n / s^2_1 & 0 \\
    0 &  \bb{Q} / s^2_{2}
\end{pmatrix}. 
\end{align*}
The latent variables are also conjugate and can be sampled from $[w_i \given \mathit{\Psi}] \stackrel{ind}\sim PG(1, h_i^\top \mathit{\Psi} )$ for $i=1, \cdots, n$, where $h_i$ is the $i$th row of $\bb H$ and $h_i^\top \mathit{\Psi} = {\varphi_i}=\psi_{1i} + \psi_{2i} $.

\paragraph{Full conditional distributions of $\sigma^2_1, \sigma^2_2$ and $s_1^2, s_2^2$.}
The full conditional for the two sets of variance parameters are
\begin{align*}
    [\sigma_1^2 &\given \bb v_1] \sim IG(n/2 + c, \bb v_1^\top \bb v_1 + d) \textspace{and}{ 2mm}
    [\sigma_2^2 \given \bb v_2] \sim IG(n/2 + c, \bb v_2^\top \bb{Q} \bb v_2 + d) \\
    [s_1^2 &\given \bb \psi_1] \sim IG(n/2 + a, \bb \psi_1^\top \bb \psi_1 + b) \textspace{and}{ 2mm} [s_2^2 \given \bb \psi_2] \sim IG(n/2 + a, \bb \psi_2^\top \bb{Q} \bb \psi_2 + b).
\end{align*}
Note that these sets of variance parameters, $(\sigma^2_1, \sigma^2_2)$ and $(s_1^2, s_2^2)$, are not necessarily on the same scale.

\begin{figure}[ht]
\centering
\includegraphics[width=0.9\textwidth]{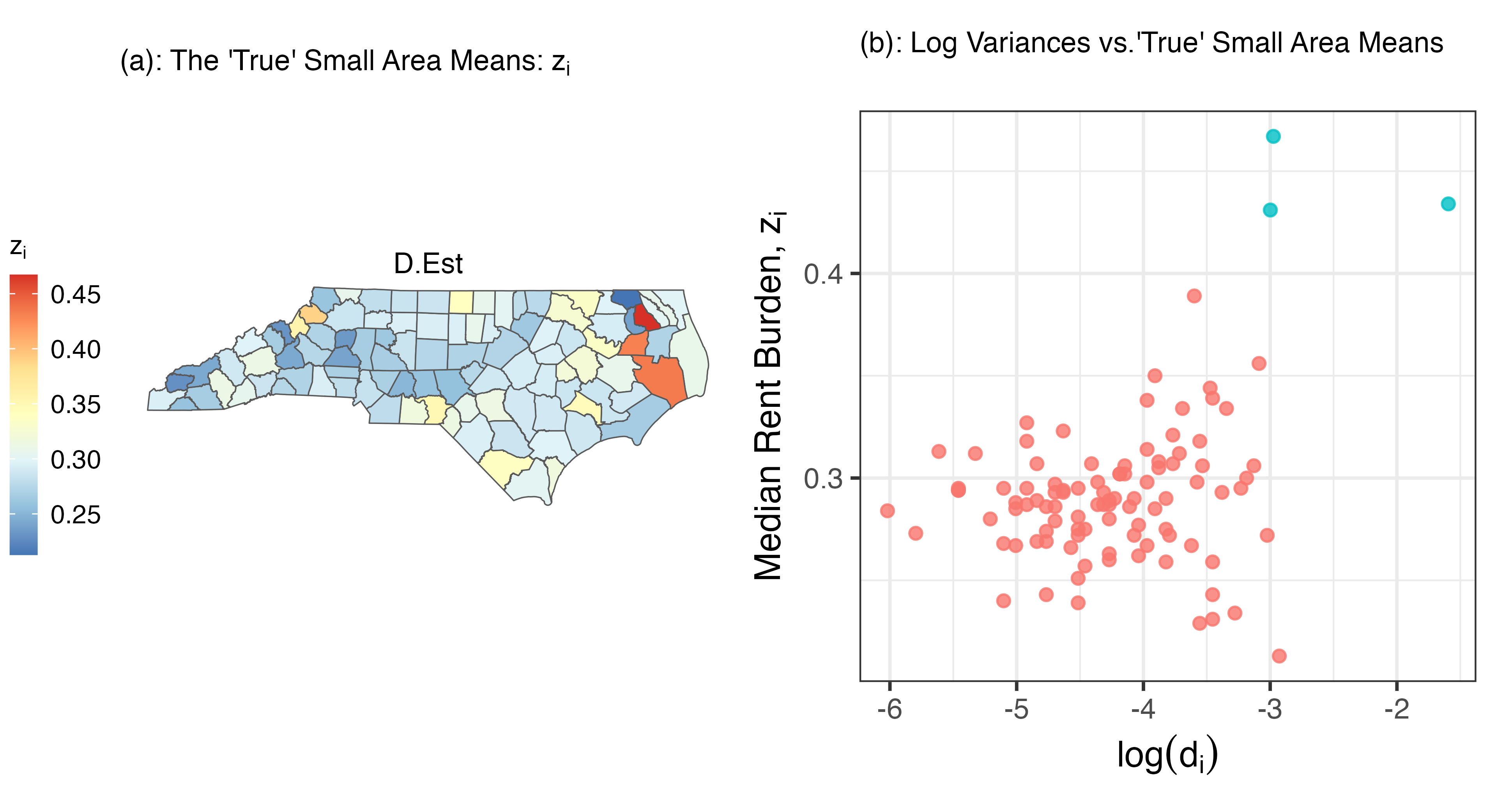}
\caption{\label{fig:z}(a) A map of the direct estimates $z_i$ of rent burden which we treat as the true small area means for our simulation study. (b) A scatterplot of the log variances $log(d_i)$ and $z_i$.  The counties with rent burdens above 40\% are highlighted in blue.}
\end{figure}

\section{Empirical Simulation Study} \label{sec:sim_study}

\subsection{Simulation Description}
In constructing our simulation study, we aim to generate data that behave similarly to what might be observed in practice. Rather than generating data synthetically from a model, we take direct estimates from an existing survey dataset and add noise based on the reported survey variances to generate data. We took this approach to preserve many of the characteristics associated with the real data when creating synthetic datasets.  The setup is similar to what is done in \citet{bradley_multivariate_2015}, \citet{bradley_computationally_2018}, and \citet{janicki_bayesian_2022}.

Specifically, the data for this simulation study was based on the publicly available American Community Survey (ACS) 5-year estimates from 2015--2019  \citep{us_census_bureau_american_2020}.  The code used to conduct this simulation study were written in R \citep{r_core_team_r_2023} using publicly available packages. The R package \texttt{tidycensus} v1.4.4 \citep{walker_tidycensus_2024} was used to download ACS data. The TIGER shapefiles for the counties were downloaded using \texttt{tigris} v2.0.3 \citep{walker_tigris_2023}. 

Let $z_i$ be the observed direct estimate of median rent as a percentage of household income (median rent burden) for the $i$th county in North Carolina. Note that there are $n=100$ counties in total. Also, let $d_i$ be the reported sampling variance associated with $z_i$. In this simulation study, we treat the $\bb z=(z_1, \cdots, z_n)^\top$ as the \textit{truth}, or the unobserved true small area means of interest. Figure  \ref{fig:z} shows the spatial distribution of the $\bb z$ and a scatterplot of $log(d_i)$ and $z_i$.  The map \ref{fig:z}(a) shows that the counties with the highest median rent burden are located fairly close together in the eastern portion of the state, providing some evidence of spatial correlation. It is notable that the counties with the lowest and highest rent burdens (Gates and Perquimans counties)  are adjacent to each other in the north eastern part of the state. The scatterplot \ref{fig:z}(b) shows that the counties with highest rent burdens also have some of the largest survey variances. Given the skew caused by these high burden counties, the robustness of the various estimators will be tested. 

In this simulation study, we run $G=100$ simulations, each with a different synthetic dataset that is created by perturbing $\bb z$ with noise distribution that uses the reported survey variances. Specifically, we generate 
$$y_i^{(g)} \stackrel{ind}{\sim} N(\theta_i = log(z_i), d_i/z_i^2)$$
for $i=1, \cdots, n$ and $g=1, \cdots, 100$, where $i$ is the index for a given county and $g$ is the index for the simulation. As rent burden can be skewed, with a lower bound at zero, we take a log transformation to better meet the Gaussian assumption. Note that since the sampling variance  for the log transformed rent burden is not known, we use the delta method to estimate this quantity.  

In each simulation, we use the synthetic dataset $\bb{y}^{(g)} = \set{ y_1^{(g)}, \cdots, y_n^{(g)} }$  and $\bb{D} = diag\{d_i/z_i^2 \}_{i=1}^{n=100}$ to predict $\bb z$. We compare the predictive performance of the SSD model with 4 other methods: 3 model-based methods as well as $\bb{y}^{(g)}$, which acts as the direct survey estimate for this synthetic dataset. The 3 other model-based estimates we compared were the independent Fay-Herriot (FH) model, the Datta-Mandal spike-and-slab model (DM), and a Fay-Herriot model with Besag-York-Mollié spatial random effects (BYM). 

The model-based estimates used $\set{\bb{y}^{(g)}, \bb{D}}$ in addition to covariates $\bb{X}$  to fit the model.  The matrix $\bb{X}$ contained ACS estimates for rates of college graduates, residents receiving public assistance income or Supplemental Nutrition Assistance Program in the past 12 months, residents who own a car, the poverty rate, racial makeup of residents (White, Black, Native, Asian), and the rate of Hispanic residents. 

The model-based approaches were all fit using Markov Chain Monte Carlo (MCMC). Each model used the same MCMC sample size ($S=2000$) with varying amounts of burn-in; the non-spatial models used 9000 iterations while the spatial models used 2000 iterations for burn-in. A randomized set of simulations were checked and lack of convergence was not detected based on a thorough visual inspection trace plots for all parameters and all models. 

The standard improper prior $\pi(\bb \beta, \sigma^2_{FH}) \propto 1$ was used for the FH. The DM used an improper prior $\pi(\bb \beta) \propto 1$, in addition to an Empirical-Bayes like prior $\sigma^2_{DM} \sim IG(3, 2 \bar{d})$, where $\bar{d} = \tfrac{1}{n}, \textstyle \sum_{i=1}^n d_i$, was recommended by the authors \citep{datta_small_2015} and acts as a scaling mechanism. The BYM model used $\pi(\bb \beta) \propto 1$ and an non-informative $IG(c, c)$ prior for $\sigma^2_1, \sigma^2_2$ with $c=5 \times 10^{-05}$. Finally, the SSD model which, again, was fit after scaling the data, used a non-informative $\bb \beta \sim N_j(\bb 0, k^2 I_j)$ with $k^2=100^2$, an $IG(5, 5)$ prior for $\sigma^2_1, \sigma^2_2$, and $IG(5, 10)$ for the logit-variance parameters $s^2_1, s^2_2$. Note that both the BYM and SSD models scaled the ICAR precision matrix. All of the models used non-informative priors for the fixed effects. Note that the random effect variance priors are not easily comparable across models. This is due to some models having one random effect term while the spatial models have two, as well as the SSD model using scaled data. 

\subsection{Assessment}
We compare the predictive capability of the various methods for $G=100$ simulations. For a given county $i$ and dataset $g$,  let $\hat{z}_i^{(g)}$ denote the point estimate (posterior mean for the model-based estimates) of the log rent burden and $\set{{\hat{l}_i}^{(g)}, {\hat{h}_i}^{(g)}}$ be the lower and upper endpoints of the $90 \%$ credible interval ($\alpha=0.1)$. We compare the performances of the various methods using mean squared error (MSE), coverage rate, and interval score, averaged across the $G=100$ simulations. We also compare the absolute bias of the estimators produced by each method.  The coverage rate and interval scores are not computed for the direct estimates as they do not produce credible intervals. 

The average of MSE serves as an overall measure of performance, balancing both bias and variance of the estimator. It is given by 
$$\mbox{Avg MSE} = \tfrac{1}{G} \sum_{g=1}^G  \frac{1}{n} \sum_{i=1}^n (\hat{z}^{(g)}_i-z_i)^2.$$
The average Coverage Rate indicates how well the $90\%$ credible intervals perform in terms of capturing the true small area means across areas and simulations. This is given by 
$$\mbox{Avg Coverage Rate} = \tfrac{1}{G} \sum_{g=1}^G \frac{1}{n} \sum_{i=1}^n \bb{I} \set{\hat{l}_i^{(g)} < z_i} \cdot \bb{I} \set{z_i< \hat{u}_i^{(g)}}.$$
The interval score is a more comprehensive way to assess interval estimates, as discussed by \citet{gneiting_strictly_2007}. The interval score penalizes for length of the interval and missed coverage on both the upper and lower endpoints. The average interval score for the simulation study is given by 
$$\mbox{Avg Interval Score}= \tfrac{1}{G} \sum_{g=1}^G  \frac{1}{n} \sum_{i=1}^n (\hat{u}_i^{(g)} - \hat{l}_i^{(g)}) + \frac{2}{\alpha} (\hat{l}_i^{(g)} - z_i) \bb{I} \set{\hat{l}_i^{(g)} > z_i} +   \frac{2}{\alpha} (z_i-\hat{u}_i^{(g)}) \bb{I}\set{z_i > \hat{u}_i^{(g)}}$$
where $\alpha$ corresponds to the $(1-\alpha) \times 100\%$ credible interval ($\alpha=0.1$ in this case). 

Finally, we also compare the average absolute bias for each estimator, given by 
$$\mbox{Avg Absolute Bias} = \frac{1}{n} \sum_{i=1}^n \bigg |  z_i - \tfrac{1}{G} \sum_{g=1}^G \hat{z}^{(g)}_i \bigg |.$$

\begin{table}[H]
\caption{\label{tab:metrics}A comparison of estimates from various methods in our empirical simulation study, which used ACS median rent burden data from North Carolina. The simulation study contained 100 simulation iterations. The Direct Estimate along with four model-based estimates were compared: independent Fay-Herriot model, the Datta-Mandal model, Fay-Herriot model with BYM effects, and the proposed SSD model. For the models, posterior means were used as point estimates and $90\%$ Credible Intervals (Cr. Int.) were used for the interval estimates. }
\centering
\begin{tabular}{l|l|l|l|l}
\hline  
Estimator& MSE&  90\% Cr. Int.& Interval Score& Absolute Bias\\
              &     &  Coverage Rate&                for 90\% Cr. Int.&    \\
\hline  
Direct Estimate& $12.3 \times 10^{-4}$& -& -& $\bb{0.0021}$\\
BYM& $6.9 \times 10^{-4}$& 0.831& 0.1150& 0.0111\\
 Fay-Herriot& $6.8 \times 10^{-4}$& 0.836& 0.1085& 0.0108\\
 Datta-Mandal& $6.5 \times 10^{-4}$& 0.793& 0.0961& 0.0110\\
 SSD Model& $\bb{5.3 \times 10^{-4}}$& $\bb{0.896}$& $\bb{0.0762}$& 0.0086\\
 \end{tabular}
\end{table}

\subsection{Results}

\begin{figure}
\centering
\includegraphics[width=0.7\textwidth]{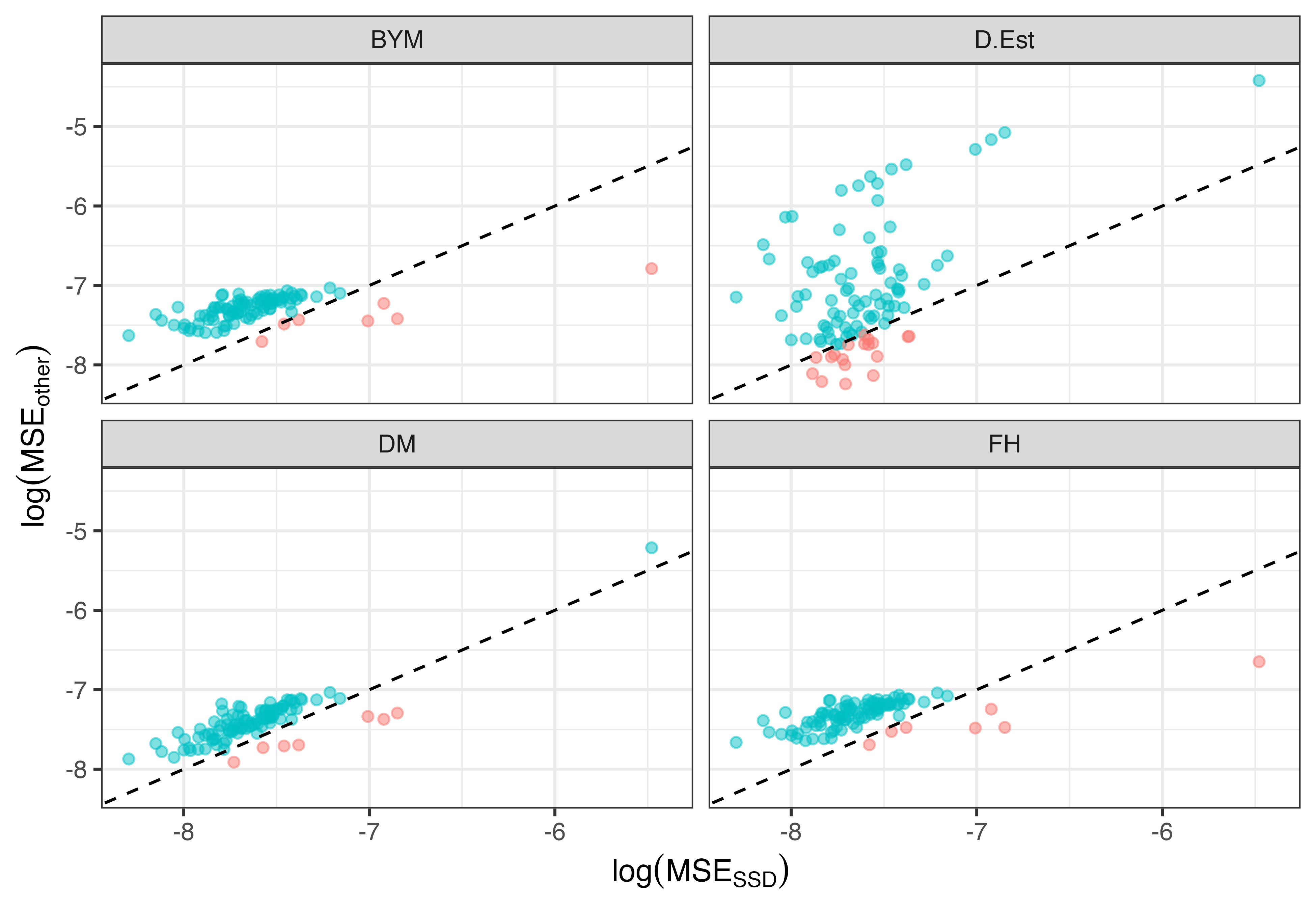}
\caption{\label{fig:mse}Comparison of the Log MSE of the estimates from the SSD model against alternative methods for each of the 100 simulations. Points above the dotted line indicate simulations where the SSD model had lower MSE while the points below the dotted line correspond to simulations where the SSD model had higher MSE. }
\end{figure}

Table~\ref{tab:metrics} summarizes the results from the simulation study.  Comparing the average MSE, the SSD model outperforms the direct estimate by 57\%, the BYM by 23\%, the FH by 22\%, and the DM by 18\%. Note that the two best performing methods used shrinkage priors for the random effects.  Figure \ref{fig:mse} compares the log MSE of the estimates of the SSD model against those from the other methods for the 100 simulations. We see that the SSD model consistently outperforms the other methods across simulations and can be used to greatly improve MSE of the small area estimates. Moreover, among the model-based approaches, the SSD model achieves the lowest bias by a 22\% margin compared to the FH (the model with the second lowest bias).  
 
 In terms of coverage rate, the SSD model is the only model whose 90\% credible intervals actually contains the truth, $\bb z$, about 90\% of the time with a coverage rate of 89.6\%.  The BYM, FH, and DM all exhibit varying degrees of undercoverage. This superior coverage is achieved without excessively large intervals, as indicated by the interval score; the SSD model has a lower interval score than the DM by 21\% (the model with the second lowest score). 

\begin{figure}
\centering
\includegraphics[width=0.9\textwidth]{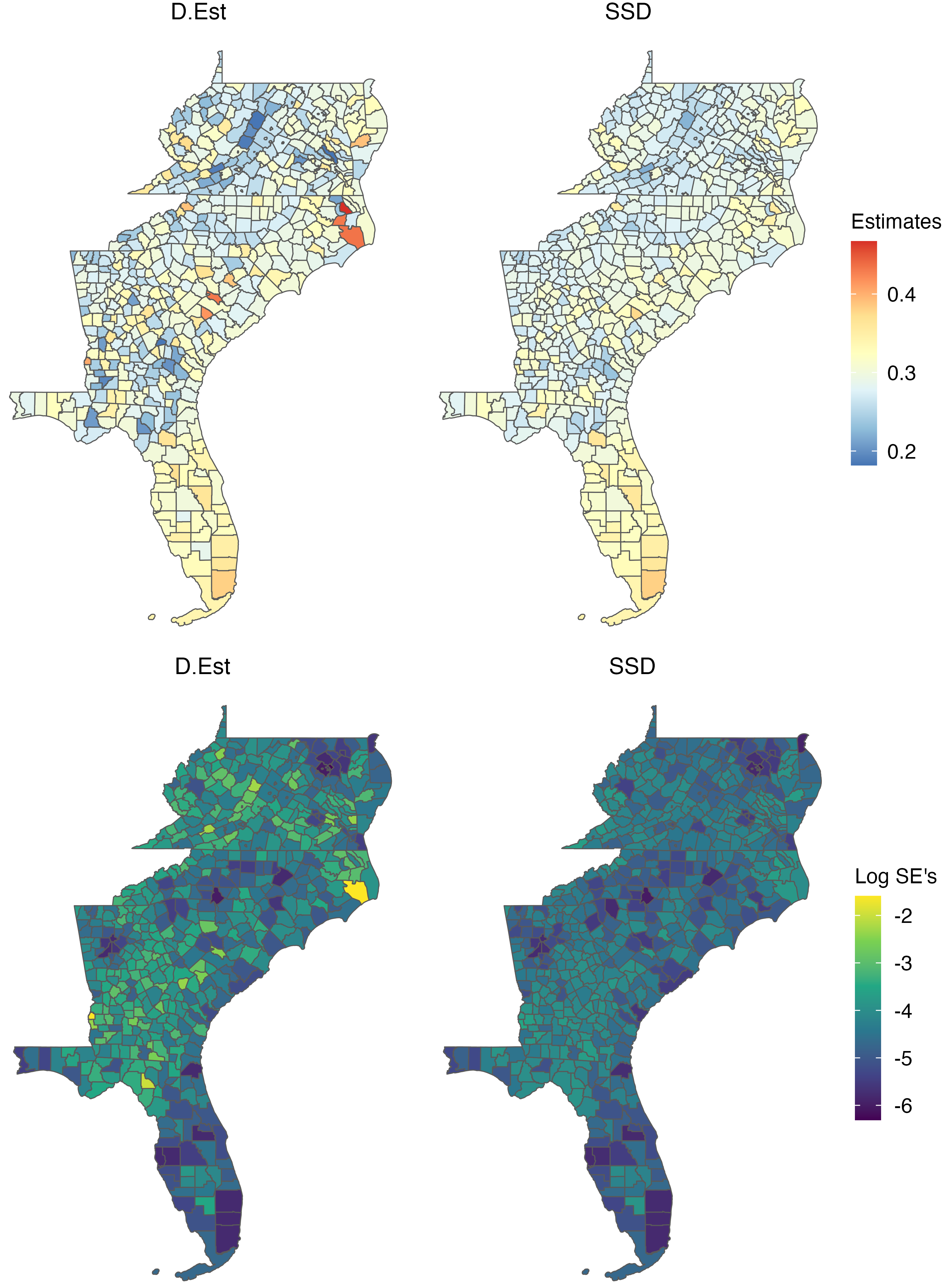}
\caption{\label{fig:estim_and_se} Comparison of the estimates and standard errors for county-level median rent burden using the direct estimates (D.Est) and the SSD model. The study region is the South Atlantic Census Division, which consists of 588 counties from D.C. and 8 states.}
\end{figure}

\section{Estimates of Median Rent Burden for the South Atlantic Census Division } \label{sec:data_analysis}

In this section, we fit the SSD model in order to estimate median rent burden in the United States South Atlantic Census Division, using 2015--2019 5-year ACS data \citep{us_census_bureau_american_2020}. The South Atlantic Census Division comprises of $8$ states plus the District of Columbia. There are $n=588$ counties in total for this Census Division.  We consider the same covariates used in Section \ref{sec:sim_study}. We used the same priors from the simulation study as well. 

The MCMC algorithm for the SSD model was run for $4000$ iterations with the first $1500$ discarded as burn-in. No lack of convergence was detected. All of the code used to conduct this analysis was written in R \citep{r_core_team_r_2023}.  Again, R packages \texttt{tidycensus} v1.4.4 \citep{walker_tidycensus_2024} and \texttt{tigris} v2.0.3 \citep{walker_tigris_2023} were used to download the necessary data. 

The maps in first row of Figure \ref{fig:estim_and_se}  compare the estimates from our proposed SSD model with the direct estimates. We see that both estimators exhibit the same overall spatial pattern, but there is a smoothing effect through the use of the model-based estimates, as expected. The impact of the smoothing is noticeable near the boundary between West Virginia and Virginia, in central South Carolina, as well as eastern North Carolina. The maps in the second row of Figure \ref{fig:estim_and_se} compare the log standard errors (SE) of the direct estimates with the log standard errors (posterior standard deviations) from the SSD model. We can see a reduction in the log SE from the SSD throughout the division. The reduction is especially noticeable in West Virginia,  Western and Eastern ends of North Carolina, Georgia, and Northern Florida.  

\begin{figure}[h]
\centering
\includegraphics[width=0.9\textwidth]{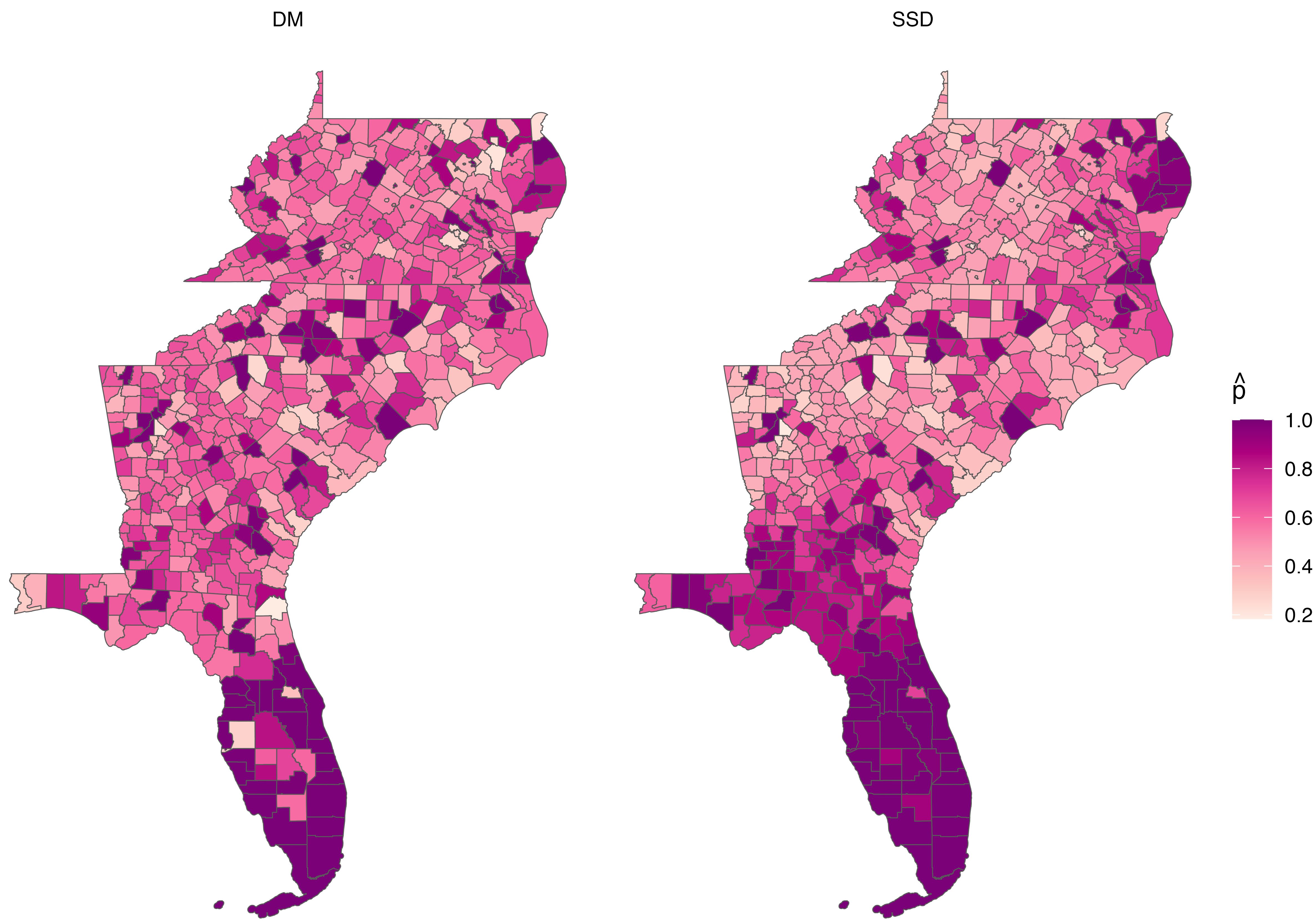}
\caption{\label{fig:iprob} Comparison of the posterior mean of the random effect selection probabilities between the DM (left) and the SSD model (right).  If the selection probability for a given county is high, the data indicates the need for inclusion of a random effect for that county.   }
\end{figure}

Finally, we fit the Datta-Mandal (DM) model to estimate median rent burden for comparison. A Monte-Carlo Simulation of Geary's C test with 1000 iterations was performed on the posterior means of the random effects from the DM model. The resulting p-value was below 0.001. This indicates that there is a strong spatial dependence in the random effects of the DM, despite the model's assumption of independence.  Figure \ref{fig:iprob} shows the posterior means of the random effect selection probabilities of the DM and SSD models.  We can see in Figure \ref{fig:iprob} that the selection probabilities of the DM model also have a strong spatial pattern. Most counties in central and southern Florida, for example, have selection probabilities close to 1.  The spatial pattern shown by the selection probabilities in the DM model is directly modeled in the SSD approach. Corespondingly, we see that the SSD selection probabilities exhibit a smoother spatial pattern. This  analysis, as well as the empirical simulation study, suggest that incorporating spatial dependence in both the random effects and the shrinkage/selection probabilities may be useful for improving estimates in certain settings. 

\section{Discussion}\label{sec:conclusion}

In this work, we develop a new small area estimation model combining both shrinkage of random effects and spatial modeling. The proposed Bayesian model incorporates spatial dependence through priors on both the random effects and the selection probabilities of the spike-and-slab prior, extending the approach by \citet{datta_small_2015}. Inference of our model can be carried out with a computationally efficient Gibbs sampler by using Pólya-Gamma data augmentation \citep{polson_bayesian_2013}.

Using data from the American Community Survey, we conducted an empirical simulation study and data analysis, both centered around estimation of median rent burden, a policy-relevant statistic. In the simulation study, we showed that the new model can produce far more accurate point and interval estimates, compared to standard approaches (direct survey estimates and the independent Fay-Herriot) and approaches that use shrinkage or spatial priors alone (Datta-Mandal and the Besag-York-Mollié). We also demonstrated the ability of the SSD model to reduce uncertainty in the data analysis compared to the direct estimate.  Both the simulation study and the data analysis illustrate the benefits of incorporating spatial dependence for both the random effects and the latent selection process. 

High computational burden is an issue for many spatial models. The Gibbs sampler for fitting our proposed model requires two inversions of the covariance matrix for every MCMC iteration: one for the blocked fixed and random effects and the other for the blocked logit parameters (see Section \ref{sec:post_inference}). These matrix inversions can make fitting the model time-intensive for applications with a large number of small areas. Future work involves development of a more efficient sampler or use of approximations to help speed up the computation for massive datasets.

The simulation study and data analysis in this article both focused on estimating county-level median rent burden using ACS data. However, the SSD model is of general interest and applicable to many other Small Area Estimation scenarios where the covariates explain the variable of interest well in many areas and the area-level random effects are expected to exhibit spatial shrinkage and dependence. The random effect priors used for the SSD model could also be adapted in a straightforward manner to spatial models in other domains such as disease mapping. 

\section*{Acknowledgements}

This research was partially supported by the U.S. National Science Foundation (NSF) under NSF Grant NCSE-2215169. This article is released to inform interested parties of ongoing research and to encourage discussion. The views expressed on statistical issues are those of the author and not those of the NSF.


\bibliographystyle{rss}
\bibliography{sources_1st_paper}

\begin{thebibliography}{46}
\expandafter\ifx\csname natexlab\endcsname\relax\def\natexlab#1{#1}\fi
\expandafter\ifx\csname url\endcsname\relax
  \def\url#1{\texttt{#1}}\fi
\expandafter\ifx\csname urlprefix\endcsname\relax\def\urlprefix{URL: }\fi

\bibitem[{Besag(1974)}]{besag_spatial_1974}
Besag, J. (1974) Spatial {Interaction} and the {Statistical} {Analysis} of {Lattice} {Systems}.
\newblock \textit{Journal of the Royal Statistical Society: Series B (Methodological)}, \textbf{36}, 192--225.
\newblock \urlprefix\url{https://rss.onlinelibrary.wiley.com/doi/10.1111/j.2517-6161.1974.tb00999.x}.

\bibitem[{Besag et~al.(1991)Besag, York and Mollié}]{besag_bayesian_1991}
Besag, J., York, J. and Mollié, A. (1991) Bayesian {Image} {Restoration}, with {Two} {Applications} in {Spatial} {Statistics}.
\newblock \textit{Annals of the Institute of Statistical Mathematics}, \textbf{43}, 1--20.
\newblock \urlprefix\url{http://link.springer.com/10.1007/BF00116466}.

\bibitem[{Bradley et~al.(2015)Bradley, Holan and Wikle}]{bradley_multivariate_2015}
Bradley, J.~R., Holan, S.~H. and Wikle, C.~K. (2015) Multivariate {Spatio-temporal} {Models} for {High}-{Dimensional} {Areal} {Data} with {Application} to {Longitudinal} {Employer}-{Household} {Dynamics}.
\newblock \textit{The Annals of Applied Statistics}, \textbf{9}.
\newblock \urlprefix\url{https://projecteuclid.org/journals/annals-of-applied-statistics/volume-9/issue-4/Multivariate-spatio-temporal-models-for-high-dimensional-areal-data-with/10.1214/15-AOAS862.full}.

\bibitem[{Bradley et~al.(2018)Bradley, Holan and Wikle}]{bradley_computationally_2018}
--- (2018) Computationally {Efficient} {Multivariate} {Spatio}-{Temporal} {Models} for {High}-{Dimensional} {Count}-{Valued} {Data} (with {Discussion}).
\newblock \textit{Bayesian Analysis}, \textbf{13}, 253--310.
\newblock \urlprefix\url{https://projecteuclid.org/journals/bayesian-analysis/volume-13/issue-1/Computationally-Efficient-Multivariate-Spatio-Temporal-Models-for-High-Dimensional-Count/10.1214/17-BA1069.full}.
\newblock Publisher: International Society for Bayesian Analysis.

\bibitem[{Chakraborty et~al.(2016)Chakraborty, Datta and Mandal}]{chakraborty_two-component_2016}
Chakraborty, A., Datta, G.~S. and Mandal, A. (2016) A {Two}-{Component} {Normal} {Mixture} {Alternative} to the {Fay}-{Herriot} {Model}.
\newblock \textit{Statistics in Transition New Series}, \textbf{17}, 67--90.
\newblock \urlprefix\url{https://www.sciendo.com/article/10.21307/stattrans-2016-006}.

\bibitem[{Cromwell(2022)}]{cromwell_renters_2022}
Cromwell, M. (2022) Renters {More} {Likely} {Than} {Homeowners} to {Spend} {More} {Than} 30\% of {Income} on {Housing} in {Almost} {All} {Counties}.
\newblock \urlprefix\url{https://www.census.gov/library/stories/2022/12/housing-costs-burden.html}.
\newblock U.S. Census Bureau Website.

\bibitem[{Crump and Schuetz(2021)}]{crump_us_2021}
Crump, S. and Schuetz, J. (2021) U.{S}. {Rental} {Housing} {Markets} are {Diverse}, {Decentralized}, and {Financially stressed}.
\newblock \textit{Web {Articles}}, The Brookings Institution.
\newblock \urlprefix\url{https://www.brookings.edu/articles/us-rental-housing-markets/}.

\bibitem[{Datta et~al.(2011)Datta, Hall and Mandal}]{datta_model_2011}
Datta, G.~S., Hall, P. and Mandal, A. (2011) Model {Selection} by {Testing} for the {Presence} of {Small}-{Area} {Effects}, and {Application} to {Area}-{Level} {Data}.
\newblock \textit{Journal of the American Statistical Association}, \textbf{106}, 362--374.
\newblock \urlprefix\url{http://www.tandfonline.com/doi/abs/10.1198/jasa.2011.tm10036}.

\bibitem[{Datta and Lahiri(1995)}]{datta_robust_1995}
Datta, G.~S. and Lahiri, P. (1995) Robust {Hierarchical} {Bayes} {Estimation} of {Small} {Area} {Characteristics} in the {Presence} of {Covariates} and {Outliers}.
\newblock \textit{Journal of Multivariate Analysis}, \textbf{54}, 310--328.
\newblock \urlprefix\url{https://www.sciencedirect.com/science/article/pii/S0047259X85710597}.

\bibitem[{Datta and Mandal(2015)}]{datta_small_2015}
Datta, G.~S. and Mandal, A. (2015) Small {Area} {Estimation} with {Uncertain} {Random} {Effects}.
\newblock \textit{Journal of the American Statistical Association}, \textbf{110}, 1735--1744.
\newblock \urlprefix\url{https://www.tandfonline.com/doi/full/10.1080/01621459.2015.1016526}.

\bibitem[{Desmond(2018)}]{desmond_heavy_2018}
Desmond, M. (2018) Heavy is the {House}: {Rent} {Burden} among the {American} {Urban} {Poor}.
\newblock \textit{International Journal of Urban and Regional Research}, \textbf{42}, 160--170.
\newblock \urlprefix\url{https://onlinelibrary.wiley.com/doi/10.1111/1468-2427.12529}.

\bibitem[{Fabrizi and Trivisano(2010)}]{fabrizi_robust_2010}
Fabrizi, E. and Trivisano, C. (2010) Robust {Linear} {Mixed} {Models} for {Small} {Area} {Estimation}.
\newblock \textit{Journal of Statistical Planning and Inference}, \textbf{140}, 433--443.
\newblock \urlprefix\url{https://www.sciencedirect.com/science/article/pii/S0378375809002304}.

\bibitem[{Fay and Herriot(1979)}]{fay_estimates_1979}
Fay, R.~E. and Herriot, R.~A. (1979) Estimates of {Income} for {Small} {Places}: {An} {Application} of {James}-{Stein} {Procedures} to {Census} {Data}.
\newblock \textit{Journal of the American Statistical Association}, \textbf{74}, 269--277.
\newblock \urlprefix\url{http://www.tandfonline.com/doi/abs/10.1080/01621459.1979.10482505}.

\bibitem[{Freni-Sterrantino et~al.(2017)Freni-Sterrantino, Ventrucci and Rue}]{freni-sterrantino_note_2017}
Freni-Sterrantino, A., Ventrucci, M. and Rue, H. (2017) A {Note} on {Intrinsic} {Conditional} {Autoregressive} {Models} for {Disconnected} {Graphs}.
\newblock \urlprefix\url{http://arxiv.org/abs/1705.04854}.
\newblock ArXiv:1705.04854 [stat].

\bibitem[{Ghosh et~al.(2022)Ghosh, Ghosh, Maples and Tang}]{ghosh_multivariate_2022}
Ghosh, T., Ghosh, M., Maples, J.~J. and Tang, X. (2022) Multivariate {Global}-{Local} {Priors} for {Small} {Area} {Estimation}.
\newblock \textit{Stats}, \textbf{5}, 673--688.
\newblock \urlprefix\url{https://www.mdpi.com/2571-905X/5/3/40}.

\bibitem[{Gneiting and Raftery(2007)}]{gneiting_strictly_2007}
Gneiting, T. and Raftery, A.~E. (2007) Strictly {Proper} {Scoring} {Rules}, {Prediction}, and {Estimation}.
\newblock \textit{Journal of the American Statistical Association}, \textbf{102}, 359--378.
\newblock \urlprefix\url{http://www.tandfonline.com/doi/abs/10.1198/016214506000001437}.

\bibitem[{Horvitz and Thompson(1952)}]{horvitz_generalization_1952}
Horvitz, D.~G. and Thompson, D.~J. (1952) A {Generalization} of {Sampling} {Without} {Replacement} from a {Finite} {Universe}.
\newblock \textit{Journal of the American Statistical Association}, \textbf{47}, 663--685.
\newblock \urlprefix\url{http://www.tandfonline.com/doi/abs/10.1080/01621459.1952.10483446}.

\bibitem[{Janicki et~al.(2022)Janicki, Raim, Holan and Maples}]{janicki_bayesian_2022}
Janicki, R., Raim, A.~M., Holan, S.~H. and Maples, J.~J. (2022) Bayesian {Nonparametric} {Multivariate} {Spatial} {Mixture} {Mixed} {Effects} {Models} with {Application} to {American} {Community} {Survey} {Special} {Tabulations}.
\newblock \textit{The Annals of Applied Statistics}, \textbf{16}, 144--168.
\newblock \urlprefix\url{https://projecteuclid.org/journals/annals-of-applied-statistics/volume-16/issue-1/Bayesian-nonparametric-multivariate-spatial-mixture-mixed-effects-models-with-application/10.1214/21-AOAS1494.full}.
\newblock Publisher: Institute of Mathematical Statistics.

\bibitem[{Jiang and Rao(2020)}]{jiang_robust_2020}
Jiang, J. and Rao, J.~S. (2020) Robust {Small} {Area} {Estimation}: an {Overview}.
\newblock \textit{Annual Review of Statistics and Its Application}, \textbf{7}, 337--360.
\newblock \urlprefix\url{https://www.annualreviews.org/doi/10.1146/annurev-statistics-031219-041212}.

\bibitem[{{Joint Center for Housing Studies of Harvard University}(2024)}]{joint_center_for_housing_studies_of_harvard_university_americas_2024}
{Joint Center for Housing Studies of Harvard University} (2024) America's {Rental} {Housing} 2024.
\newblock \urlprefix\url{https://www.jchs.harvard.edu/sites/default/files/reports/files/Harvard_JCHS_Americas_Rental_Housing_2024.pdf}.

\bibitem[{Leroux et~al.(2000)Leroux, Lei and Breslow}]{leroux_estimation_2000}
Leroux, B.~G., Lei, X. and Breslow, N. (2000) Estimation of {Disease} {Rates} in {Small} {Areas}: a {New} {Mixed} {Model} for {Spatial} {Dependence}.
\newblock In \textit{Statistical {Models} in {Epidemiology}, the {Environment}, and {Clinical} {Trials}} (eds. M.~E. Halloran and D.~Berry), 179--191. New York, NY: Springer.

\bibitem[{Park and Casella(2008)}]{park_bayesian_2008}
Park, T. and Casella, G. (2008) The {Bayesian} {Lasso}.
\newblock \textit{Journal of the American Statistical Association}, \textbf{103}, 681--686.
\newblock \urlprefix\url{https://www.tandfonline.com/doi/full/10.1198/016214508000000337}.

\bibitem[{Parker et~al.(2023{\natexlab{a}})Parker, Holan and Janicki}]{parker_conjugate_2023}
Parker, P.~A., Holan, S.~H. and Janicki, R. (2023{\natexlab{a}}) Conjugate {Modeling} {Approaches} for {Small} {Area} {Estimation} with {Heteroscedastic} {Structure}.
\newblock \textit{Journal of Survey Statistics and Methodology}, smad002.
\newblock \urlprefix\url{https://academic.oup.com/jssam/advance-article/doi/10.1093/jssam/smad002/7058158}.

\bibitem[{Parker et~al.(2023{\natexlab{b}})Parker, Janicki and Holan}]{parker_comprehensive_2023}
Parker, P.~A., Janicki, R. and Holan, S.~H. (2023{\natexlab{b}}) A {Comprehensive} {Overview} of {Unit}-{Level} {Modeling} of {Survey} {Data} for {Small} {Area} {Estimation} {Under} {Informative} {Sampling}.
\newblock \textit{Journal of Survey Statistics and Methodology}, \textbf{11}, 829--857.
\newblock \urlprefix\url{https://academic.oup.com/jssam/article/11/4/829/7198195}.

\bibitem[{Petrucci and Salvati(2006)}]{petrucci_small_2006}
Petrucci, A. and Salvati, N. (2006) {Small} {Area} {Estimation} for {Spatial} {Correlation} in {Watershed} {Erosion} {Assessment}.
\newblock \textit{Journal of Agricultural, Biological, and Environmental Statistics}, \textbf{11}, 169.
\newblock \urlprefix\url{https://doi.org/10.1198/108571106X110531}.

\bibitem[{Polson et~al.(2013)Polson, Scott and Windle}]{polson_bayesian_2013}
Polson, N.~G., Scott, J.~G. and Windle, J. (2013) Bayesian {Inference} for {Logistic} {Models} using {Pólya}–{Gamma} {Latent} {Variables}.
\newblock \textit{Journal of the American Statistical Association}, \textbf{108}, 1339--1349.
\newblock \urlprefix\url{https://doi.org/10.1080/01621459.2013.829001}.
\newblock Publisher: Taylor \& Francis \_eprint: https://doi.org/10.1080/01621459.2013.829001.

\bibitem[{Porter et~al.(2015)Porter, Wikle and Holan}]{porter_small_2015}
Porter, A.~T., Wikle, C.~K. and Holan, S.~H. (2015) Small {Area} {Estimation} via {Multivariate} {Fay}–{Herriot} {Models} with {Latent} {Spatial} {Dependence}.
\newblock \textit{Australian \& New Zealand Journal of Statistics}, \textbf{57}, 15--29.
\newblock \urlprefix\url{https://onlinelibrary.wiley.com/doi/abs/10.1111/anzs.12101}.
\newblock \_eprint: https://onlinelibrary.wiley.com/doi/pdf/10.1111/anzs.12101.

\bibitem[{Quigley(2008)}]{quigley_housing_2008}
Quigley, J.~M. (2008) Housing {Policy} in the {United} {States}.
\newblock \urlprefix\url{https://escholarship.org/uc/item/89p9r7w9}.

\bibitem[{{R Core Team}(2023)}]{r_core_team_r_2023}
{R Core Team} (2023) \textit{R: {A} {Language} and {Environment} for {Statistical} {Computing}}.
\newblock Vienna, Austria: R Foundation for Statistical Computing.
\newblock \urlprefix\url{https://www.R-project.org/}.

\bibitem[{Reich and Staicu(2021)}]{tadesse_bayesian_2021}
Reich, B.~J. and Staicu, A.-M. (2021) Bayesian {Variable} {Selection} in {Spatial} {Regression} {Models}.
\newblock In \textit{Handbook of {Bayesian} {Variable} {Selection}}, 251--270. Boca Raton: Chapman and Hall/CRC, 1 edn.
\newblock \urlprefix\url{https://www.taylorfrancis.com/books/9781003089018/chapters/10.1201/9781003089018-11}.

\bibitem[{Riebler et~al.(2016)Riebler, Sørbye, Simpson and Rue}]{riebler_intuitive_2016}
Riebler, A., Sørbye, S.~H., Simpson, D. and Rue, H. (2016) An {Intuitive} {Bayesian} {Spatial} {Model} for {Disease} {Mapping} that {Accounts} for {Scaling}.
\newblock \textit{Statistical Methods in Medical Research}, \textbf{25}, 1145--1165.
\newblock \urlprefix\url{https://doi.org/10.1177/0962280216660421}.
\newblock Publisher: SAGE Publications Ltd STM.

\bibitem[{Schmid and Münnich(2014)}]{schmid_spatial_2014}
Schmid, T. and Münnich, R.~T. (2014) Spatial {Robust} {Small} {Area} {Estimation}.
\newblock \textit{Statistical Papers}, \textbf{55}, 653--670.
\newblock \urlprefix\url{https://doi.org/10.1007/s00362-013-0517-y}.

\bibitem[{Singh et~al.(2005)Singh, Shukla and Kundu}]{singh_spatio-temporal_2005}
Singh, B.~B., Shukla, G.~K. and Kundu, D. (2005) Spatio-{Temporal} {Models} in {Small} {Area} {Estimation}.
\newblock \textit{Survey Methodology}.

\bibitem[{Sugasawa et~al.(2017)Sugasawa, Tamae and Kubokawa}]{sugasawa_bayesian_2017}
Sugasawa, S., Tamae, H. and Kubokawa, T. (2017) Bayesian {Estimators} for {Small} {Area} {Models} {Shrinking} {Both} {Means} and {Variances}.
\newblock \textit{Scandinavian Journal of Statistics}, \textbf{44}, 150--167.
\newblock \urlprefix\url{https://onlinelibrary.wiley.com/doi/10.1111/sjos.12246}.

\bibitem[{Sørbye and Rue(2014)}]{sorbye_scaling_2014}
Sørbye, S.~H. and Rue, H. (2014) Scaling {Intrinsic} {Gaussian} {Markov} {Random} {Field} {Priors} in {Spatial} {Modelling}.
\newblock \textit{Spatial Statistics}, \textbf{8}, 39--51.
\newblock \urlprefix\url{https://www.sciencedirect.com/science/article/pii/S2211675313000407}.

\bibitem[{Tang and Ghosh(2023)}]{tang_global-local_2023}
Tang, X. and Ghosh, M. (2023) Global-{Local} {Priors} for {Spatial} {Small} {Area} {Estimation}.
\newblock \textit{Calcutta Statistical Association Bulletin}, \textbf{75}, 141--154.
\newblock \urlprefix\url{https://doi.org/10.1177/00080683231186378}.
\newblock \_eprint: https://doi.org/10.1177/00080683231186378.

\bibitem[{Tang et~al.(2018)Tang, Ghosh, Ha and Sedransk}]{tang_modeling_2018}
Tang, X., Ghosh, M., Ha, N.~S. and Sedransk, J. (2018) Modeling {Random} {Effects} using {Global}–{Local} {Shrinkage} {Priors} in {Small} {Area} {Estimation}.
\newblock \textit{Journal of the American Statistical Association}, \textbf{113}, 1476--1489.
\newblock \urlprefix\url{https://www.tandfonline.com/doi/full/10.1080/01621459.2017.1419135}.

\bibitem[{{U.S. Bureau of Labor Statistics}(1914)}]{us_bureau_of_labor_statistics_consumer_1914}
{U.S. Bureau of Labor Statistics} (1914) Consumer {Price} {Index} for {All} {Urban} {Consumers}: {Rent} of {Primary} {Residence} in {U}.{S}. {City} {Average}.
\newblock \urlprefix\url{https://fred.stlouisfed.org/series/CUUR0000SEHA}.
\newblock Publisher: FRED, Federal Reserve Bank of St. Louis.

\bibitem[{{U.S. Census Bureau}(2020)}]{us_census_bureau_american_2020}
{U.S. Census Bureau} (2020) American {Community} {Survey} 2015-2019 5-{Year} {Estimates}.
\newblock Accessed via tidycensus package.

\bibitem[{{U.S. Census Bureau}(2023)}]{us_census_bureau_american_2023}
--- (2023) American {Community} {Survey} 2022 1-{Year} {Estimates}.
\newblock \urlprefix\url{https://www.census.gov/programs-surveys/acs/data.html}.

\bibitem[{Wakefield(2007)}]{wakefield_disease_2007}
Wakefield, J. (2007) Disease {Mapping} and {Spatial} {Regression} with {Count} {Data}.
\newblock \textit{Biostatistics (Oxford, England)}, \textbf{8}, 158--183.

\bibitem[{Walker(2023)}]{walker_tigris_2023}
Walker, K. (2023) \textit{tigris: {Load} {Census} {TIGER}/{Line} {Shapefiles}}.
\newblock \urlprefix\url{https://CRAN.R-project.org/package=tigris}.

\bibitem[{Walker and Herman(2024)}]{walker_tidycensus_2024}
Walker, K. and Herman, M. (2024) \textit{tidycensus: {Load} {US} {Census} {Boundary} and {Attribute} {Data} as 'tidyverse' and 'sf'-{Ready} {Data} {Frames}}.
\newblock \urlprefix\url{https://walker-data.com/tidycensus/}.

\bibitem[{You(2016)}]{you_hierarchical_2016}
You, Y. (2016) Hierarchical {Bayes} {Sampling} {Variance} {Modeling} for {Small} {Area} {Estimation} {Based} on {Area} {Level} {Models} with {Applications}.
\newblock \textit{Methodology branch working paper, ICCSMD-2016-03-E, Statistics Canada, Ottawa, Canada}.

\bibitem[{You(2021)}]{you_small_2021}
--- (2021) {Small} {Area} {Estimation} {Using} {Fay}-{Herriot} {Area} {Level} {Model} with {Sampling} {Variance} {Smoothing} and {Modeling}.
\newblock \textit{Survey Methodology}, \textbf{47}, 361--370.
\newblock \urlprefix\url{https://www150.statcan.gc.ca/n1/en/pub/12-001-x/2021002/article/00007-eng.pdf?st=eb5cMHvz}.

\bibitem[{Zhou and You(2008)}]{zhou_hierarchical_2008}
Zhou, Q.~M. and You, Y. (2008) Hierarchical {Bayes} {Small} {Area} {Estimation} for the {Canadian} {Community} {Health} {Survey}.
\newblock \textit{Survey Methodology}, \textbf{37}, 25--37.

\end{thebibliography}
\end{document}